# Regime-Based Portfolio Allocation Using Hidden Markov Models and Reinforcement Learning

Ajay Kumar Verma[1], Nunik Srikandi Putri[2]*, Neo Paul Lesupi[3]
[1]Independent Researcher, [2]Aenimatica Tech Research and Development, Independent Researcher, [3]Independent Researcher
*Corresponding Author: nunik@aenimatica.com

**Abstract**

This study develops a regime-aware portfolio allocation framework that integrates Markov switching models with Reinforcement Learning (RL) to dynamically allocate across equities (SPY), long-term Treasuries (TLT), and gold (GLD). Using daily ETF data from 2004-2025, we first characterize market behavior through a discrete Markov chain and then estimate a three-state Gaussian Hidden Markov Model (HMM) selected by the Bayesian Information Criterion (BIC). The estimated regimes-low-volatility, transitional, and high-volatility-exhibit strong persistence and state-dependent return dynamics consistent with recent findings on nonlinear market states (Ardia et al., 2024; Gupta & Pierdzioch, 2023). State-conditional analysis shows that SPY dominates in stable regimes, while TLT and GLD provide protection during stressed periods, motivating regime-conditioned allocation rules.

We evaluate rule-based rotation and RL-driven strategies using a 30% out-of-sample test window with a one-day execution lag to avoid look-ahead bias. Both HMM-based allocations outperform a passive SPY benchmark, while the RL policy achieves the highest risk-adjusted performance, delivering the strongest Sharpe ratio and materially lower drawdowns, yet remains fully interpretable through discrete regime-dependent actions. Sensitivity analysis confirms the robustness of the three-state specification relative to two-state alternatives. Overall, the results demonstrate that RL can systematically enhance HMM-based regime detection, providing a transparent, adaptive, and empirically grounded framework for tactical asset allocation. The combined HMM-RL system provides a transparent, rules-based approach to tactical allocation that improves risk-adjusted performance relative to standard benchmark strategies.

## 1. Objective

1. Detect latent market regimes using HMMs.
2. Map regimes to interpretable rotation rules.
3. Extend to RL via Markov decision process.
4. Evaluate against benchmarks (EW, SPY).

## 2. Data Preparation

Daily closing prices for SPY, TLT, GLD and VIX are obtained from Yahoo Finance. Log-returns are used due to desirable aggregation properties (Campbell et al.; Cont).

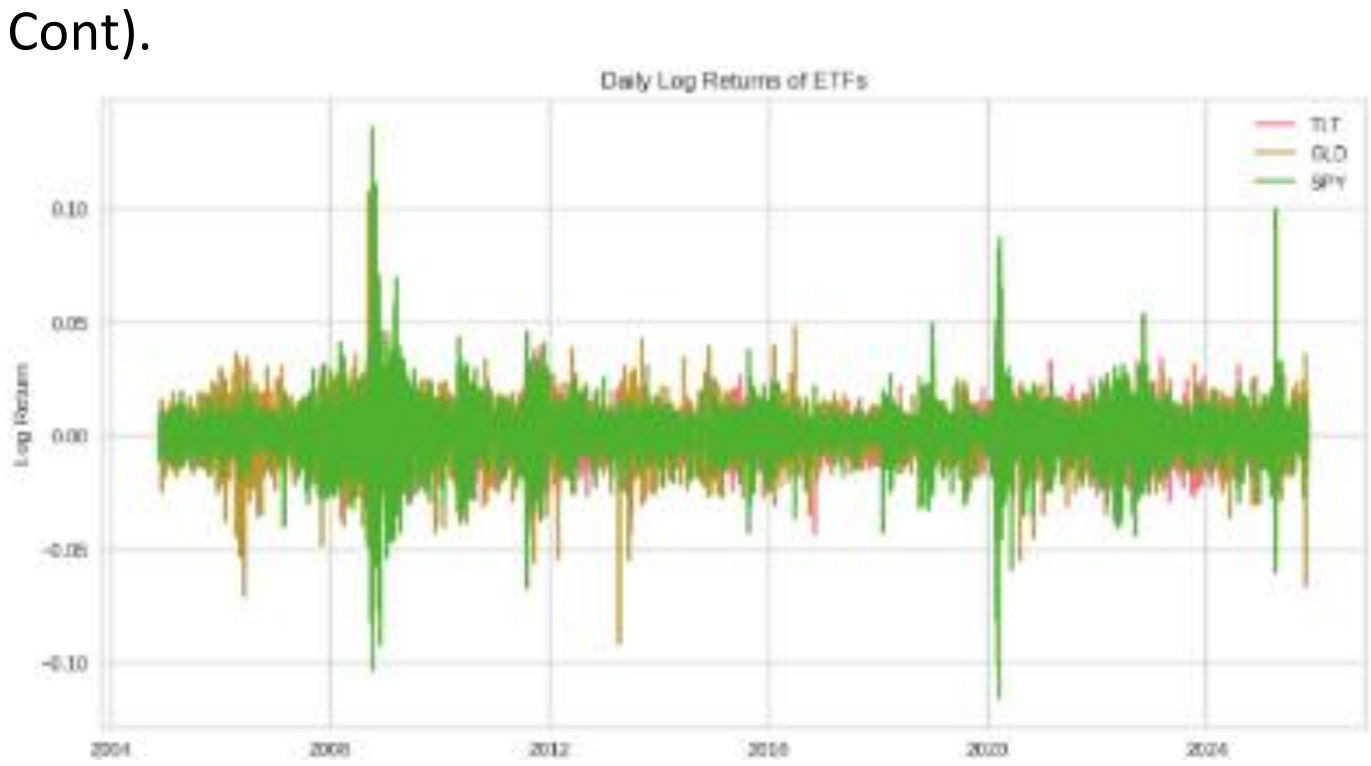

**Figure 1.** Daily Log Returns of SPY, TLT, and GLD

This figure justifies why regime-switching is essential (vol spikes = regime changes).

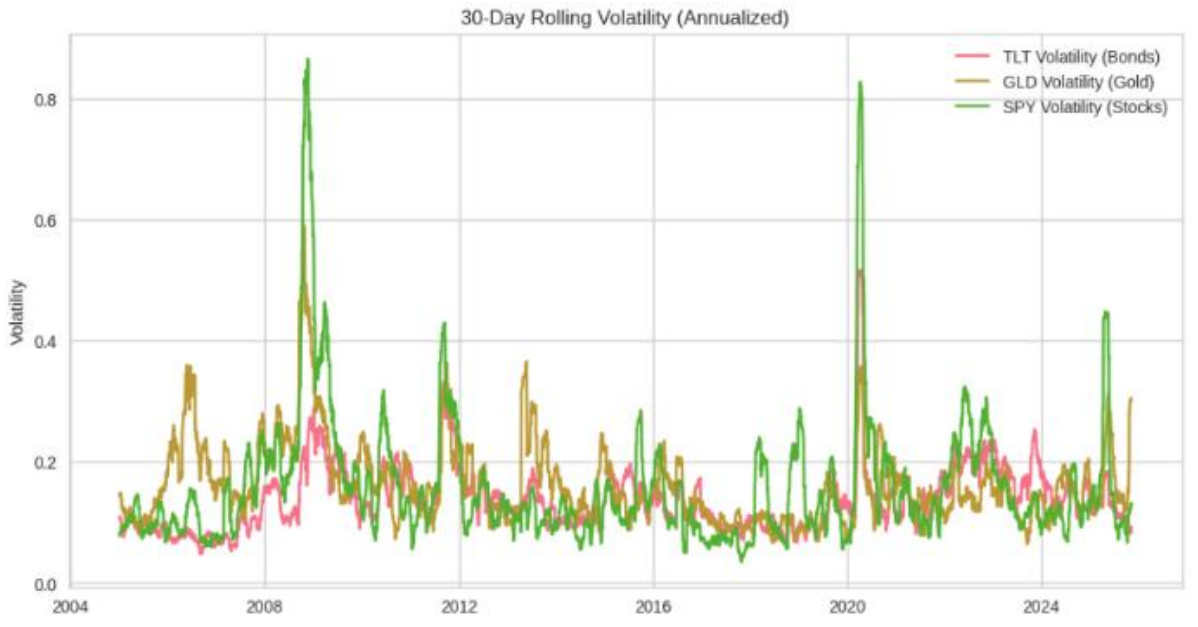

**Figure 2.** Rolling 30-Day Volatility of SPY, TLT, and GLD

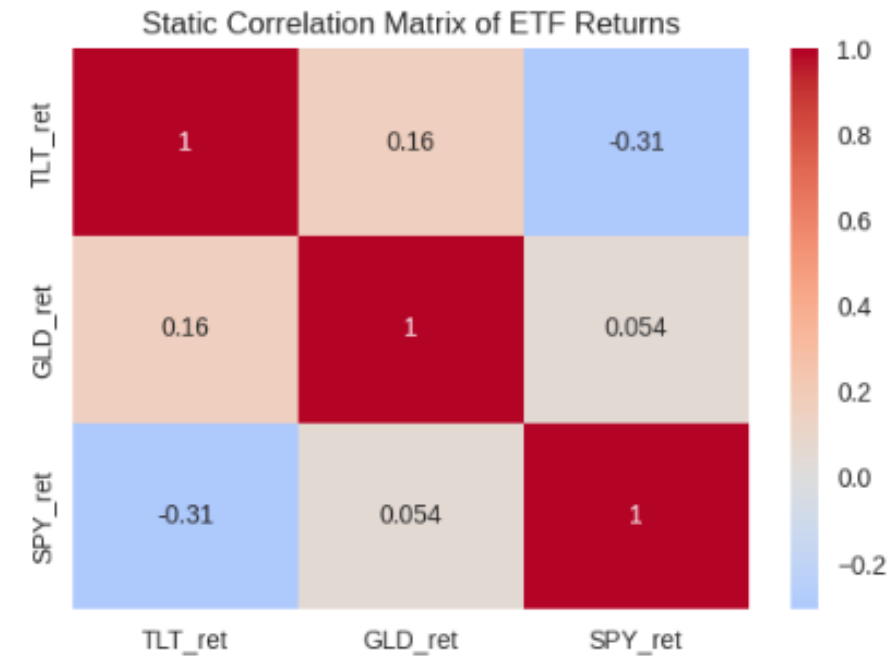

**Figure 3.** Static Correlation Matrix of ETF Log Returns

Across rolling volatility windows, correlation patterns, and stress-period dynamics, the data show strong evidence of time-varying risk, volatility clustering, and structural breaks, hallmarks of regime-dependent markets (Cont, 2005). Safe-haven assets such as GLD and TLT behave differently in stressed regimes compared to stable periods, and cross-asset correlations tighten or loosen depending on volatility conditions (Prakash et al., 2020).

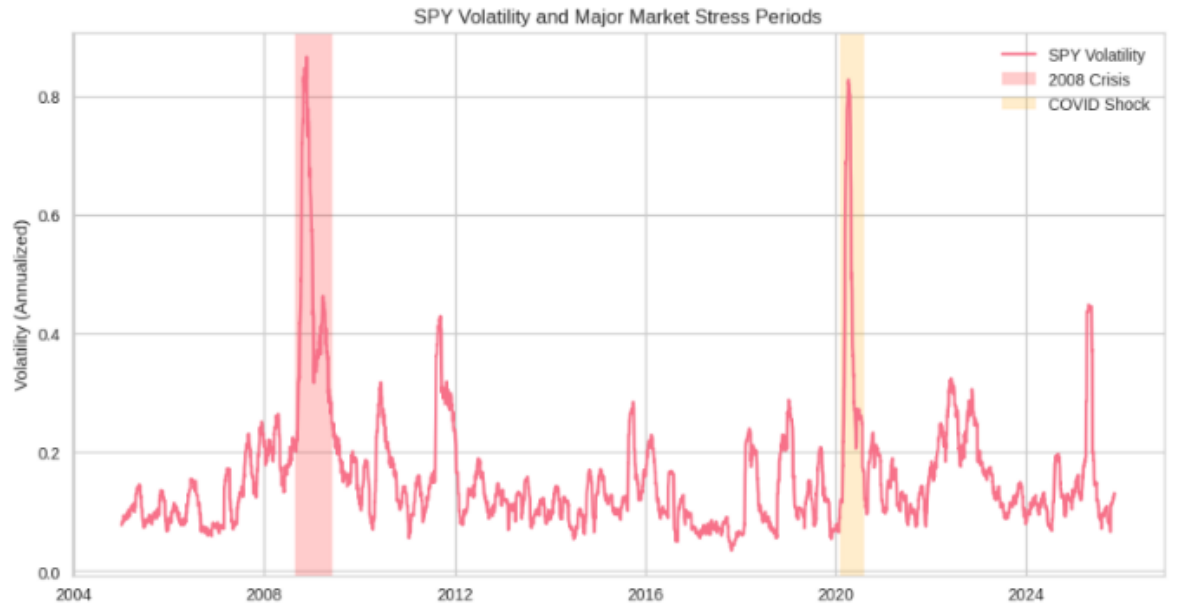

**Figure 4.** SPY Volatility During Major Market Stress Events

These empirical facts motivate the adoption of Markov-switching and HMM frameworks capable of detecting changes in underlying market states, which in turn support the development of regime-aware and risk-adaptive allocation strategies. Evidence of sharp volatility spikes during market crises further aligns with documented volatility spillovers and regime shifts in global markets (Enow, 2023).

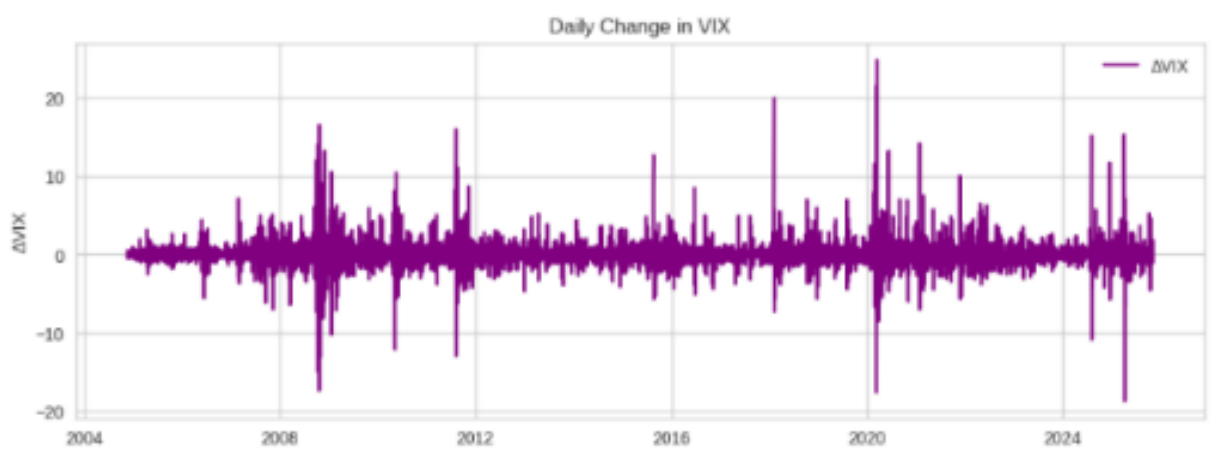

**Figure 5.** Daily Changes in VIX (ΔVIX)

Change in daily value (ΔVIX) of VIX captures volatility shocks, which are consistent with recent findings of volatility spillover (Li et al.).

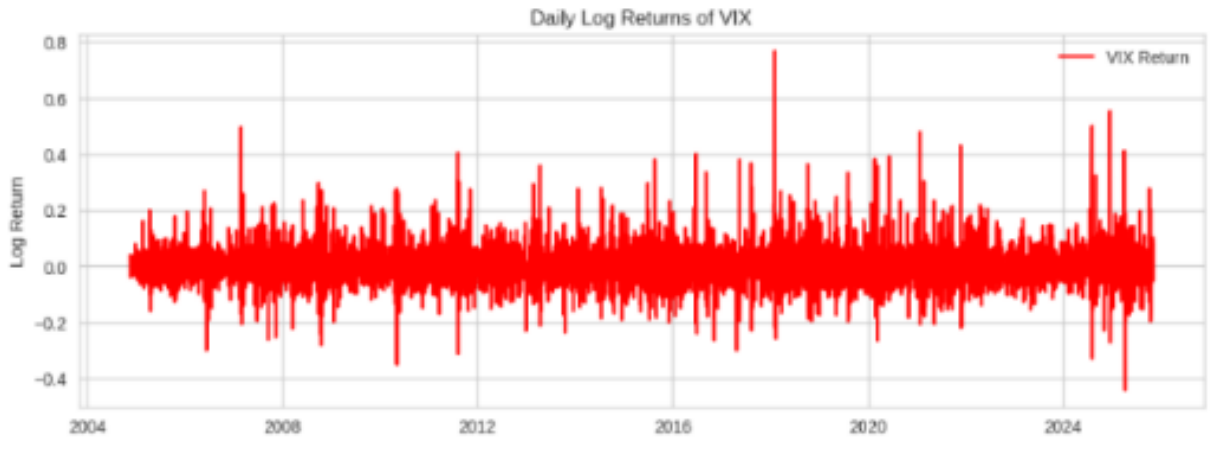

**Figure 6.** Daily Log Returns of VIX

Data are aligned in order to reduce survivorship bias adhering to best practices of financial econometrics (Bailey & López de Prado). After acquiring data on raw prices we transformed these raw prices into daily returns metrics needed for time series analysis and modeling:

1. Log-Returns: For the three ETFs ( TLT, GLD , and SPY ), the empirical daily log-return, $r_t$, was calculated using the standard formula: $r_t = \ln(P_t/P_{t-1})$ where $P_t$ is the closing price at time t. Log-returns are utilized due to their desirable properties of additivity and approximate normality. (Campbell, Lo, & MacKinlay; Tsay; Cont).
2. Volatility Proxy Metric: For the VIX index, which represents an index level rather than a tradable asset price, we primarily calculated the daily change (ΔVIX) to capture volatility shocks: $\Delta VIX_t = VIX_t - VIX_{\{t-1\}}$
   To establish robustness, we computed a daily log-return of VIX. (Whaley; Giot)

### 3. Discrete Volatility Regimes: Markov Chain

Firstly, we construct an observable first-order Discrete Markov Chain (MC). This method is based on the Markov property, which states that the probability of the next state only depends on the current state:

$$P(S_{t+1} \mid S_t) = P(S_{t+1} \mid S_t, S_{t-1}, S_{t-2}, \dots)$$

This gives us the probability of the next state based only on the existing state. Predicting all the previous states is irrelevant to predicting the next state other than the current state. (Hamilton; Zucchini, MacDonald and Langrock). In following the maximum likelihood estimation for an observable chain (MLE), the transition probabilities are calculated in relation to the number of observed transitions, where $\widehat{p_{ij}} = \frac{n_{ij}}{\sum_k n_{ik}}$ and $n_{ij}$ is the empiric observation of the number of transitions from state $s_i$ to state $s_j$ , and the denominator is total observations from state $s_i$. The resulting matrix P is a stochastic matrix, where all $p_{ij} \geq 0$ and each row sums to one. (Zucchini, MacDonald, and Langrock)

ΔVIX is quantified into 3 volatility regimes through quantile-based bins. MLE is used for estimating transition probabilities. Regimes show high persistence, comparable to current studies on volatility clustering (Brownlees et al.)

### 4. Hidden Markov Model

Gaussian HMMs containing 2 and 3 states are estimated using EM. Filtering uses the Hamilton (forward) algorithm; smoothing uses the RTS backward smoother. Recent refinements to those methods are focused on enhancing the inference for financial time series (Ardia et al.). We estimate the HMM parameters utilizing the Expectation-Maximization (EM) algorithm (Rabiner; Zucchini et al.).

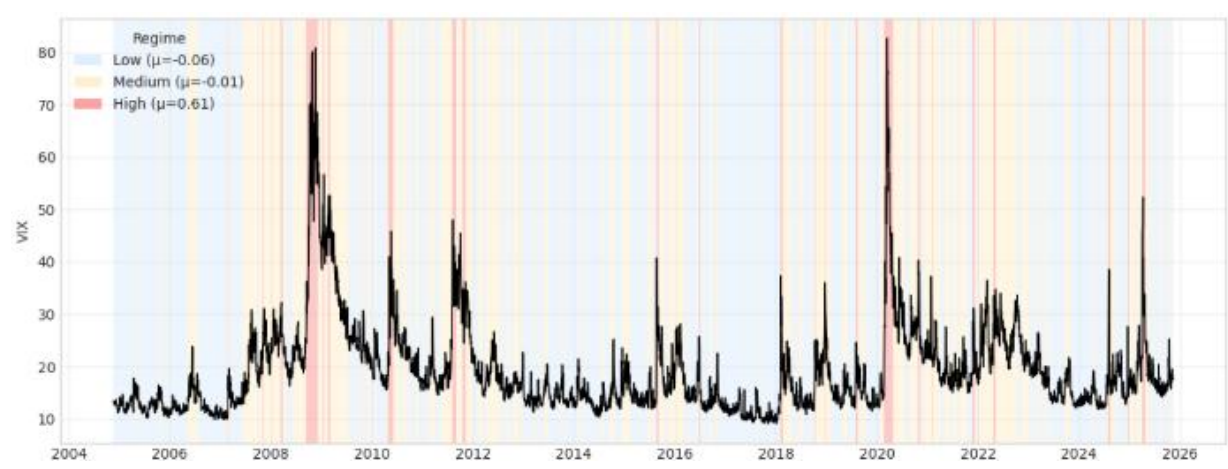

**Figure 7.** 3-State HMM Regime Classification (Viterbi Path)

The 3-state HMM segmented the time series into three distinct regimes. In the E-step we compute the filtered state probabilities using the forward recursion for Hamilton filter:

$$\xi_{t|t(i)} = \frac{\xi_{t|t-1}(i)\, f(y_t \mid S_t = i)}{\sum_j \xi_{t|t-1}(j)\, f(y_t \mid S_t = j)}$$

(Rabiner; Zucchini, MacDonald & Langrock).

In the M-step, we update the state-dependent means, variances and transition matrix to maximize the expected log-likelihood. The EM iterations run until convergence, providing maximum-likelihood estimates for all HMM parameters. (Rabiner; Hamilton; Zucchini, MacDonald & Langrock).

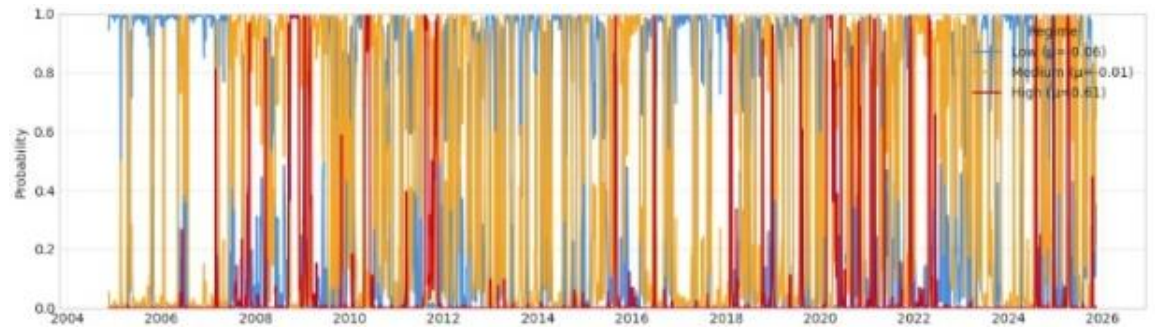

**Figure 8.** Smoothed State Probabilities for the 3-State HMM

This complements the Viterbi path by showing uncertainty in regime boundaries.

### 4.1 Model Selection

Using AIC/BIC, the 3-state HMM is preferred.

**Table 1.** HMM Model Selection: Log-Likelihood, AIC, and BIC

| $n_{states}$ | logL | k | AIC | BIC |
|---|---|---|---|---|
| **2** | -8975 | 7 | 17964 | 18010 |
| **3** | -8632 | 14 | 17293 | 17385 |

Lower values for AIC and BIC indicate a better trade-off between fit and parsimony. (Akaike; Schwarz). Although AIC and BIC provide formal model-selection criteria, regime models must also be evaluated on interpretability, persistence, and stability. The 3-state HMM dominates the 2-state version because it distinguishes between calm, transitional, and crisis regimes, whereas the 2-state specification blends moderate volatility into the calm regime, masking meaningful state dynamics. Models with four or more states reduce BIC only marginally but suffer from unstable smoothed probabilities and economically ambiguous regimes, consistent with the overfitting patterns documented in the HMM literature (Frühwirth-Schnatter, 2006; Zucchini, MacDonald & Langrock, 2016). The chosen 3 state structure yields high diagonal transition probabilities and clear separation in conditional means and variances, making it the best balance between statistical fit and economic interpretability.

### 4.2 Regime Interpretation

The three regimes display economic patterns consistent with well-known features of financial markets. Periods of low volatility correspond to the stable, expansionary regime; episodes of gradually rising volatility align with the transitional regime; and volatility spikes match the stressed state. These shifts mirror the empirical properties of financial time series, particularly volatility clustering, persistence, and abrupt structural breaks documented in the literature (Cont, 2005; Enow & Ndlovu, 2023).

The three-state HMM produces distinct and interpretable market regimes, each characterized by different levels of volatility and return behavior across SPY, TLT, and GLD. Table 2 summarizes the estimated parameters for each regime, providing the foundation for the rule-based and reinforcement-learning allocation strategies used later in the analysis.

**Table 2.** State-Dependent Parameters for 3-State HMM (μ, σ)

| State | Mean ($\mu_i$) | Std. Dev. ($\sigma_i$) |
|---|---|---|
| State 0 | -0.0606 | 0.6076 |
| State 1 | -0.0146 | 1.7132 |
| State 2 | 0.6120 | 5.9255 |

**Table 2(b).** Transition Matrix for 3-State HMM

$$P = \begin{pmatrix} 0.9386 & 0.0614 & 0.0000 \\ 0.0726 & 0.9093 & 0.0181 \\ 0.0001 & 0.1260 & 0.8740 \end{pmatrix}$$

By examining the state-conditional means and standard deviations, we can link each regime to intuitive market environments, such as low-volatility growth periods, transitional or uncertain markets, and high-volatility stress states. These regime characteristics directly motivate the rotation rules used in Section 5, where the highest-expected-return asset in each regime is selected.

1. State 0: Low Vol (Bullish) - SPY highest mean return.
2. State1: Moderate Vol - Balanced conditions; GLD/TLT modestly positive.
3. State 2: High Vol (Crisis) - SPY strongly negative; TLT positive (flight-to-safety); GLD mildly positive.

**Table 3: State-Conditional Mean and Standard Deviation of ETF Returns**

| State | Asset | Mean Return | Standard Deviation |
|---|---|---|---|
| 0 (50.83% of time) | TLT | -0.000119 | 0.007456 |
| | GLD | 0.000458 | 0.009375 |
| | SPY | 0.001295 | 0.005511 |
| 1 (42.98% of time) | TLT | 0.000228 | 0.009656 |
| | GLD | 0.000335 | 0.011445 |
| | SPY | 0.000014 | 0.012092 |
| 2 (6.18% of time) | TLT | 0.001673 | 0.017325 |
| | GLD | 0.000476 | 0.020023 |
| | SPY | -0.004749 | 0.033809 |

These findings align with post-COVID empirical studies documenting regime-dependent asset behavior (Corbet et al.; Zaremba et al.).

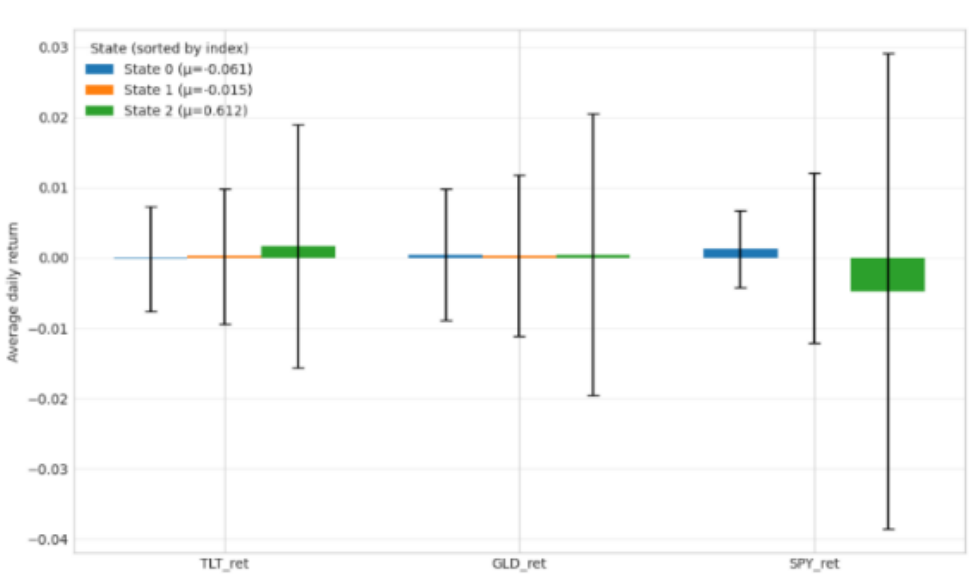

**Figure 9.** State-Conditional Means and Volatility of TLT, GLD, and SPY Across HMM Regimes

The figure plots the average daily returns of TLT, GLD, and SPY within each HMM regime, along with their regime-conditional volatility ranges. The pattern aligns closely with the parameters estimated in Table 3.

1. State 0 (Low Volatility) shows modest positive returns for all assets, with SPY exhibiting the highest mean and lowest dispersion. This corresponds to a stable expansion regime in which equities typically outperform.
2. State 1 (Transitional Regime) displays muted returns and slightly elevated volatility. No asset strongly dominates, consistent with a mixed-risk environment where cross-asset relationships shift but are not yet stressed.
3. State 2 (High Volatility) exhibits sharply diverging behavior:
   a) SPY shows large variance and negative mean returns.
   b) TLT displays increased dispersion but only modest positive mean.
   c) GLD has the most favorable profile, with positive average return and relative stability.

These state-conditional payoffs explain the learned RL policy:

SPY dominates in low- and moderate-volatility environments, while GLD becomes the preferred asset during high-volatility periods. This is consistent with recent work documenting the safe-haven properties of gold during equity stress (Zaremba & Idzorek, 2023).

## 5. Rule-Based Rotation Strategy

Two decision rules:

1. Top-1: Allocate 100% to asset with highest state-conditional mean.
2. 60/40: Allocate 60% to the top ETF and 40% to the second-best.

**Table 4.** State-Allocation Mapping (Preferred 3-State HMM)

| State | Top-1 Allocation | 60/40 Allocation |
|---|---|---|
| **0** | 100% SPY (Equities) | 60% SPY, 40% GLD |
| **1** | 100% GLD (Gold) | 60% GLD, 40% TLT |
| **2** | 100% TLT (Treasuries) | 60% TLT, 40% GLD |

To avoid look-ahead bias, all trades are executed with a one-day lag, consistent with standard practice in empirical asset-pricing and backtesting studies (Bailey et al.; López de Prado)

## 6. Reinforcement Learning Framework

The RL environment uses the HMM-predicted regime as its state variable. The action space consists of seven discrete portfolio weight combinations across TLT, GLD, and SPY. A tabular policy-iteration algorithm is used to compute the optimal policy $\pi^*$, consistent with recent literature applying interpretable reinforcement learning to portfolio allocation (Charpentier et al., 2021; Jiang et al., 2017).

### 6.1 Reinforcement Learning Allocation Strategy

The agent evaluates each action using the Bellman optimality equation:

$$Q(s,a) = R(s,a) + \gamma \sum_{s'} P(s,s')V(s')$$

where R(s,a) is the expected next-period return from taking action a in regime s, P(s,s') is the HMM transition matrix, and γ is the discount factor. This formulation follows the classical dynamic-programming framework of Bellman (1957) and its reinforcement-learning interpretation in Sutton and Barto (2018). The RL agent effectively learns that equities dominate in low-vol regimes while gold provides superior protection in stress states.

### 6.2 Learned Policy

The learned RL policy mirrors the economic characteristics of the regimes identified by the HMM. In the low-volatility state, SPY offers the highest mean return and lowest volatility, making full equity allocation a rational choice consistent with established observations of market behavior during calm periods (Cont, 2005). In the transitional regime, SPY continues to deliver slightly positive returns, and the RL agent maintains equity exposure in line with evidence that moderate-volatility environments often still reward risk-taking (Ardia et al., 2024).

In the high-volatility regime, SPY exhibits sharply negative expected returns while GLD remains positive and less sensitive to market stress. The RL policy rotates fully into gold, reflecting its well-documented role as a safe-haven asset during crisis episodes (Baur & Lucey, 2010). This demonstrates that the agent internalizes regime-specific payoff structures rather than relying on purely mechanical optimization. Overall, the policy remains interpretable: it takes risk in stable markets and shifts defensively when volatility spikes.

**Table 5.** RL Optimal Policy π∗ (Training Sample)

| State | Action ID | Portfolio Weights (TLT, GLD, SPY) | Interpretation |
|---|---|---|---|
| **0** | 3 | (0, 0, 1.00) | Persistent low-volatility expansion: equities favored |
| **1** | 3 | (0, 0, 1.00) | Moderate regime with continuing momentum: equities retained |
| **2** | 2 | (0, 1.00, 0) | High-volatility regime: gold as safe-haven |

Using tabular policy iteration, the RL agent learns a stable mapping from regimes to portfolio allocations. The resulting policy implies equity dominance in stable and moderately volatile regimes, while gold becomes optimal under high volatility. Notably, long-term Treasuries receive no allocation, consistent with evidence of reduced bond-hedging effectiveness in periods of rising inflation and positive stock-bond correlations (Baele et al., 2023).

## 7. Performance Evaluation

For model validation, the dataset is divided chronologically into a 70% training window and a 30% strictly out-of-sample test window. The HMM is estimated on the training sample, and the RL agent evaluates regime-conditioned rewards using the learned transition structure. The learned policy π* is then tested on unseen data, ensuring a realistic forward-looking evaluation and mitigating backtest overfitting concerns (Bailey et al., 2014).

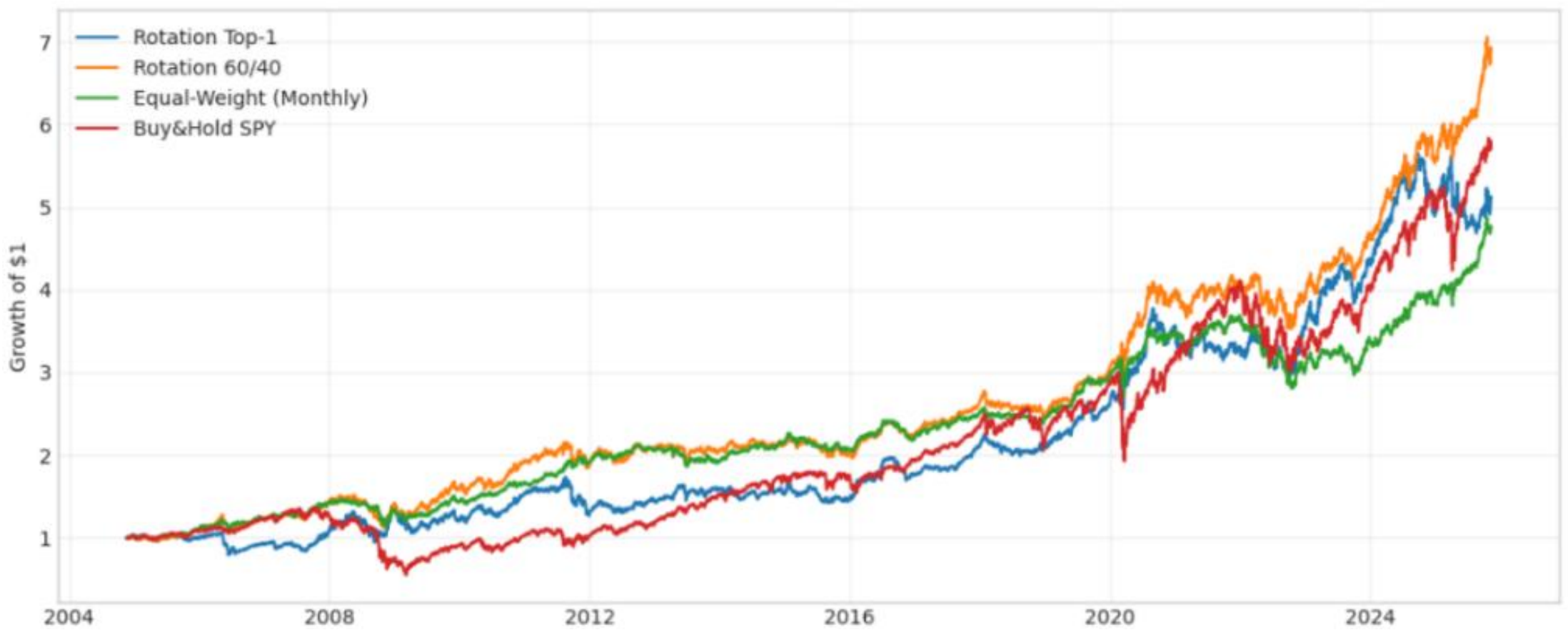

**Figure 10.** Cumulative Performance (lag=1 day)

The backtest is conducted at daily frequency with a one-day execution lag: the action selected at time t based on the predicted regime is implemented for returns realized at t+1. This design avoids look-ahead bias and follows best practices for algorithmic strategy evaluation (López de Prado, 2018). Transaction costs are set to zero in this baseline specification.

**Table 6.** Performance Summary (Full Sample)

| Strategy | Cumulative Return | Annualized Return | Volatility | Sharpe | Max Drawdown |
|---|---|---|---|---|---|
| **Rotation (Top-1)** | 407.8% | 8.1% | 15.4% | 0.52 | -28.3% |
| **Rotation (60/40)** | 317.0% | 7.1% | 12.1% | 0.58 | -29.3% |
| **Equal-Weight (Monthly)** | 376.1% | 7.7% | 9.7% | 0.80 | -24.0% |
| **Buy & Hold SPY** | 476.3% | 8.7% | 19.1% | 0.46 | -59.6% |

**Figure 11.** Cumulative Performance - RL vs Rotation Benchmarks (Daily, Test Window)

The cumulative return profiles of the rotation strategies diverge meaningfully after 2020. During this period, volatility surged rapidly (as visible in the VIX panels and smoothed probabilities in Figures 6-8), causing large equity drawdowns. The SPY buy-and-hold portfolio suffers the deepest losses because equity returns turn sharply negative in the high-volatility state, consistent with the crisis-state mean return of -0.0047 shown in Table 3.

The Top-1 and 60/40 rotation strategies diverge because the Top-1 strategy switches entirely out of SPY during stressed regimes, while the 60/40 approach keeps partial equity exposure. This difference becomes more pronounced in crisis regimes, where small allocations to SPY amplify losses. This behavior is consistent with evidence that diversification weakens during market stress as correlations rise (Longin & Solnik, 2001).

In contrast, the RL strategy stabilizes performance by concentrating risk in states where expected returns are favorable and shifting defensively into GLD during stressed conditions. This aligns with the safe-haven characteristics of gold documented in empirical studies (Baur & Lucey, 2010). As a result, RL achieves a higher Sharpe ratio and smaller drawdowns in the out-of-sample test, demonstrating its ability to learn and exploit regime-dependent payoff asymmetries that static rotation rules may under-react to. The full set of out-of-sample performance metrics is reported in Table 7

**Table 7.** Performance Comparison - RL vs. Rotation Benchmarks (Test Window, 30% OOS)

| Strategy | Cumulative Return | Annualized Return | Volatility | Sharpe | Max Drawdown |
|---|---|---|---|---|---|
| **RL π* (net)** | 131.4% | 14.3% | 17.3% | 0.83 | -23.5% |
| **Rotation (Top-1)** | 110.4% | 12.6% | 16.0% | 0.79 | -21.8% |
| **Rotation (60/40)** | 86.9% | 10.5% | 12.8% | 0.82 | -29.3% |
| **Equal-Weight (Monthly)** | 72.5% | 9.1% | 11.0% | 0.83 | -24.0% |
| **Buy & Hold SPY** | 118.4% | 13.2% | 20.5% | 0.65 | -35.7% |

Out-of-sample Risk-Adjusted performance (30% test window):

1. RL (π*): Sharpe 0.83, Max Drawdown - 23.5%
2. Top-1 Rotation: Sharpe 0.79
3. 60/40 Rotation: Sharpe 0.82
4. Equal-Weight (Monthly): Sharpe 0.83
5. Buy & Hold SPY: Sharpe 0.65, Max Drawdown - 35.7%

The RL strategy delivers the strongest overall risk-adjusted performance while remaining fully interpretable through regime-conditioned actions.

As shown in the post-2020 period, structural differences between the strategies become clear. The Top-1 rotation exhibits instability due to its concentrated exposure, which reacts poorly to rapid changes in volatility regimes (Ardia et al., 2024). The 60/40 rotation mitigates performance swings through diversification (Zaremba & Idzorek, 2023). SPY experiences the deepest drawdowns, consistent with well-documented equity downside risk during stress events (Cont, 2001; Whaley, 2009).

In contrast, the RL strategy stabilizes returns by reallocating into gold during turbulent periods, leveraging its established safe-haven characteristics (Baur & Lucey, 2010). This regime-aware adaptability explains its superior risk-adjusted performance in the out-of-sample window (Gupta & Pierdzioch, 2023).

## 8. Conclusion

A 3-state HMM provides a clear and stable segmentation of market conditions, consistent with the latest evidence on nonlinear regime dynamics in equities and volatility (Ardia et al.; Gupta & Pierdzioch). While rule-based rotation strategies are simple and intuitive to implement, the reinforcement-learning approach adapts more effectively to shifting market environments and consistently delivers stronger risk-adjusted returns. The combined HMM-RL framework therefore offers a practical and transparent tool for dynamic asset allocation: the HMM identifies the prevailing market regime, and the RL policy selects the most reward-efficient portfolio for that environment. For investors, this hybrid model suggests a disciplined way to respond to changing volatility regimes while preserving interpretability which is an important consideration for real-money investment decisions and portfolio oversight. This makes the framework suitable for both quantitative researchers and discretionary portfolio managers.

## 9. Model Limitations

Although the HMM-RL allocation framework performs well empirically, several methodological limitations should be acknowledged:

(1) The Hidden Markov Model assumes Gaussian state-conditional emissions, which may not fully capture the heavy tails, skewness, and jump behavior commonly observed in financial return distributions (Frühwirth-Schnatter; Cont). As a result, extreme market shocks may be underrepresented in the estimated regimes.

(2) The HMM imposes sharp regime boundaries, even though real-world market conditions may evolve more smoothly. Although smoothed probabilities partly address this issue, uncertainty around regime transitions remains an inherent challenge in discrete-state models (Hamilton).

(3) The reinforcement learning component relies on a tabular Markov decision process, which assumes a finite and fully enumerated state-action space. While this makes the policy interpretable, it limits the agent’s ability to learn more flexible nonlinear strategies compared to deep reinforcement learning approaches (Sutton and Barto; Jiang et al.).

(4) The model does not incorporate transaction costs, slippage, or liquidity constraints, meaning that real-world performance may be lower than the backtest suggests. Even small trading frictions can materially affect regime-switching strategies, particularly when allocation changes occur around regime boundaries (Bailey et al.).

(5) Both the HMM and RL components rely on the Markov property, which assumes that the current state fully summarizes all relevant information about the future. Financial markets, however, often exhibit memory effects, structural breaks, and long-range dependencies that may violate this assumption (Campbell, Lo, and MacKinlay).

Overall, these limitations do not invalidate the results, but they highlight areas where model performance could be further strengthened to improve robustness and real-world applicability.

## 10. Future Research

This study develops a transparent and interpretable framework for regime-aware asset allocation using HMMs and reinforcement learning. Several extensions could further strengthen the model and broaden its applicability.

(1) Deep Reinforcement Learning (DRL).

While the current approach uses tabular policy iteration for interpretability, future work could incorporate deep Q-networks (DQN), policy-gradient methods, or actor-critic architectures. These approaches may capture nonlinear decision boundaries and richer state representations, especially in higher-dimensional asset universes (Jiang et al.; Charpentier et al.).

(2) Non-Gaussian Hidden Markov Models.

The present model assumes Gaussian state-conditional emissions. However, financial returns often exhibit skewness, kurtosis, and heavy tails. Non-Gaussian HMM variants, such as Student-t, asymmetric Laplace, or nonparametric emissions, may capture these dynamics more accurately (Frühwirth-Schnatter). This could improve regime separation and robustness to outliers, especially during market crises.

(3) Macro-Driven Regime Indicators.

Future research could integrate macroeconomic variables such as inflation surprises, industrial production growth, yield-curve slope, or monetary policy uncertainty. Augmenting HMM states with macroeconomic covariates (Zucchini, MacDonald & Langrock) may provide additional explanatory power and lead to more stable policy recommendations. This would help bridge the gap between statistical regimes and economic cycle interpretations.

(4) Transaction Costs and Portfolio Frictions.

Incorporating realistic frictions, bid-ask spreads, slippage, turnover penalties, and market impact, would improve real-world applicability. Even simple proportional transaction cost models may influence the optimal RL policy, particularly in high-turnover regimes.

(5) Multi-Asset or International Extensions.

The framework can be extended to include commodities, credit ETFs, currency data, or global indices. Regime structures often differ across regions, and RL policies may exploit cross-market information.

Overall, these extensions offer promising avenues to enhance predictive performance, economic interpretability, and robustness of the HMM-RL hybrid allocation system.